**Sun-to-Earth Propagation of the 2015 June 21 Coronal Mass Ejection Revealed by Optical, EUV, and Radio Observations**

N. Gopalswamy[1*], P. Mäkelä[1,2], S. Akiyama[1,2], S. Yashiro[1,2], H. Xie[1,2], and N. Thakur[1,2]

[1]NASA Goddard Space Flight Center, Greenbelt, MD 20771, USA

[2]The Catholic University of America, Washington DC 20064, USA

* Corresponding author. Email: nat.gopalswamy@nasa.gov

**Abstract**

We investigate the propagation of the 2015 June 21 CME-driven shock as revealed by the type II bursts at metric and longer wavelengths and coronagraph observations. The CME was associated with the second largest geomagnetic storm of solar cycle 24 and a large solar energetic particle (SEP) event. The eruption consisted of two M-class flares, with the first one being confined, with no metric or interplanetary radio bursts. However, there was intense microwave burst, indicating accelerated particles injected toward the Sun. The second flare was eruptive that resulted in a halo CME. The CME was deflected primarily by an equatorial coronal hole that resulted in the modification of the intensity profile of the associated SEP event and the duration of the CME at Earth. The interplanetary type II burst was particularly intense and was visible from the corona all the way to the vicinity of the Wind spacecraft with fundamental-harmonic structure. We computed the shock speed using the type II drift rates at various heliocentric distances and obtained information on the evolution of the shock that matched coronagraph observations near the Sun and in-situ observations near Earth. The depth of the geomagnetic storm is consistent with the 1-AU speed of the CME and the magnitude of the southward component.

**1. Introduction**

Understanding solar eruptive events that produce significant impact on the heliosphere and on Earth in particular are of great interest in developing prediction schemes for space weather. By significant impact we mean the production of intense geomagnetic storms, the occurrence of sudden commencement and the acceleration of particles to very high energies. All these space-weather effects occurred following the 2015 June 21 coronal mass ejection (CME), which has been the subject of several investigations on various aspects (Baker et al. 2016; Liu et al., 2015; Reiff et al. 2016; Astafyeva et al. 2015; Piersanti et al. 2017). The event is also part of the campaign event study undertaken by SCOSTEP's Variability of the Sun and Its Terrestrial Impact (VarSITI) projects (Webb and Nitta 2017; Marubashi et al. 2017).

One of the controversies in this event is the low speed of the CME (~330 km s$^{-1}$) in the outer corona (~26 solar radii) assumed by Piersanti et al. (2017), based on measurements provided by the CORIMP automatic CME catalog. With such low speed, these authors estimated the shock arrival at Earth on June 25, while the shock arrived on June 22, about 39.5 hours after the eruption. In fact, the shock and interplanetary (IP) CME (ICME) speeds were much higher at 1 AU (>700 km s$^{-1}$). This is one of the motivations to clarify the CME kinematics near the Sun because we know that CMEs that survive into the IP medium to reach Earth are generally fast and wide (Gopalswamy et al. 2010a). We show that the shock speed is consistently high throughout the inner heliosphere and the deceleration is typical of most CMEs reported in the literature. We also use multiple methods to infer the true speed of the shock from white-light and radio data. In particular, we use IP type II radio burst data to derive the shock speed in the inner heliosphere. The CME and the shock are tightly coupled, so it is not possible to have a 330 km s$^{-1}$ CME driving a >1000 km s$^{-1}$ shock.

The solar source of the CME has not received much attention in the literature (Liu et al. 2015; Piersanti et al. 2017). Soft-X-ray observations show two M-class flares in quick succession from the same active region. We examine the spatial distribution of the flare emission to show that it is different during the two flares: the first flare (X-ray class M2.0) was confined (no mass motion), while the second one (M2.6) was eruptive and hence associated with the CME in question.

Another motivation stems from the confusion regarding the nature and extent of the IP counterpart of the 2015 June 21 CME observed at 1 AU. Some authors identified small two flux ropes (duration ~4 h each) inside an extended single ICME (Liu et al. 2015), while others have argued for a single short-duration flux rope (Marubashi et al. 2017). Detailed investigation of the coronal flux rope suggests that the flux rope size and magnetic content near the Sun are not compatible with the corresponding 1-AU properties of the ICME. We provide evidence for deflection of the CME near the Sun by a set of coronal holes that might have resulted in the shorter-than-usual size of the ICME.

There were several halo CMEs around the time of the 2015 June 21 halo CME. The space weather consequences of these CMEs were limited except for the June 21 CME, which produced the second largest geomagnetic storm in solar cycle 24 and a large SEP event. Using white-light, radio, and in-situ observations, we develop a scenario that seems to explain both the large SEP event and the magnetic superstorm.

**2. Data and Methods**

The CME data used in this paper are from the Large Angle and Spectrometric Coronagraph (LASCO, Brueckner et al. 1995) on board the Solar and Heliospheric Observatory (SOHO) mission. We use the original images as well as the measurements already available at the SOHO/LASCO CME catalog (Yashiro et al. 2004; Gopalswamy et al. 2009a). Information on the source region and the early phase of the eruption are obtained from the Atmospheric Imaging

Assembly (AIA, Lemen et al. 2012) and the Helioseismic and Magnetic Imager (HMI, Scherrer et al. 2012) on board the Solar Dynamics Observatory (SDO). We use H-alpha images from the Big Bear Solar Observatory to get the geometry of the neutral line from the filament. The Nobeyama Radioheliograph (Nakajima et al. 1994) provided the microwave flare images. We use metric radio data from the Culgoora and Hiraiso Radio spectrographs to analyze the early manifestation of the CME-driven shock. To describe the IP propagation of the CME and the shock driven by it, we use the Radio and Plasma Wave Experiment (WAVES, Bougeret et al. 1995) on board the Wind spacecraft. The WAVES data on the type II radio burst at frequencies from 13.4 MHz down to 20 kHz are obtained by the three receivers: RAD2, RAD1, and the thermal noise receiver (TNR). We obtain information on the soft X-ray flare and the SEP event from GOES data. To connect the solar source to the CME, we use the FRED technique (Flux Rope from Eruption Data, Gopalswamy et al. 2017a,b,c) that uses photospheric magnetograms in conjunction with EUV and white-light images. Finally, we use in-situ data available online from Operating Missions as a Node on the Internet (OMNI, http://omni.gsfc.nasa.gov).

*Table 1. Events around the time of the 2015 June 21 CME*

| Date & Time | Speed (km s$^{-1}$) | Flare Size & Location | Type II Range | SEP (pfu) | IP Shock[d] | ICME | Dst (nT) |
|---|---|---|---|---|---|---|---|
| 06/18 17:24 | 1305 | M3.0 N15E50 | DH-km | HiB[b] | 6/21 15:50 (S1) | No | ---- |
| 06/19 06:42 | 584 | FC  S25W00[a] | No | No | 6/22 04:52 (S2) | No | ---- |
| 06/21 02:36 | 1366 | M2.6  N12E13 | m-km | 1066 | 6/22 18:01 (S3) | 6/23 02:00[e] | -204 |
| 06/22 18:36 | 1209 | M6.5  N12W08 | m-km | HiB[c] | 6/24 13:00 (S4) | 6/25 09:30 | -86 |
| 06/25 08:36 | 1627 | M7.9  N09W42 | m-km | 22 | 6/27 02:00 (S5) | 6/28 00:30 | -55 |

[a]*Approximate centroid of a long filament channel (FC) that erupted to the east of AR 12367;* [b]*The high background is due to an earlier fast (1714 km s$^{-1}$) and wide (279º) CME at 01:25 UT from the west limb; Possible event with an intensity of ~200 pfu (particle flux unit) right after the ESP event from the June 21 CME;* [d]*The shocks S1-S4 were also identified in Liu et al. (2015);* [e]*See also Liu et al. (2015); Marubashi et al. (2017).*

## 3. Analysis and Results

**3.1 Overview of Events around the Time of the June 21 CME**

The 2015 June 21 CME originated from NOAA AR 12371 located close to the disk center (N12E13). The active region first rotated on to the disk on 2015 June 16. The first significant CME occurred on June 17 at 03:12 UT, which was of moderate speed (549 km s$^{-1}$) and width (50º). The second CME was on June 18 at 02:36 UT, also of moderate speed (784 km s$^{-1}$) and width (60º). Following this there was a series of five halo CMEs on June 18, 19, 21, 22, and 25 as listed in Table 1. All except the June 19 CME were from AR 12371; the June 19 CME originated from a filament channel eruption to the south of the disk center. There were also two episodes of large-scale disturbances in the region surrounding AR 12371 (especially to the south and west) that did not have associated CMEs probably because they were not wide enough to become halos. All halo CMEs from AR 12371 were associated with type II bursts over a wide range as noted in Table 1: metric (m), decameter-hectometric (DH) and kilometric (km) wavelengths. The CMEs were separated significantly and there was no overlap in the occurrence of type II bursts. Details of the type II bursts can be obtained from the Javascript movies available at https://cdaw.gsfc.nasa.gov/CME_list/radio/waves_type2.html.

The source of the June 18 CME was too far to the east from the Sun-Earth line, so only the western flank of the shock (S1) arrived at Wind. Similarly the June 19 CME was far to the south, so Wind passed through just the northern flank of the shock S2. Wind missed the ejecta in both cases. The type II burst associated with the June 18 CME started around 7 MHz with fundamental-harmonic structure but became clear only around 3 MHz. The burst continued to frequencies below ~500 kHz, but ended around 20 UT on June 18 well before S1 arrived at Wind. There was no type II burst associated with the June 19 CME, so the associated shock S2 was radio quiet. The type II burst associated with the June 21 CME started at metric wavelengths and continued all the way to the local plasma frequency at Wind, coincident with the arrival of S3. The type II burst associated with the June 22 CME started after the arrival of shock S3 at Wind; the burst ended below 300 kHz towards the end of June 22 and the associated shock S4 arrived at Wind on June 24 at 13 UT. Finally the type II burst associated with the June 25 CME ended around 150 kHz on the same day by 16:30 UT, while the associated shock S5 arrived at Wind at 02:00 UT on June 27. The 1-AU counterparts of CMEs on June 21, 22, and 25 were observed in-situ by Wind and resulted in geomagnetic storms with strengths of -204 nT (June 23 at 5 UT), -86 nT (June 25 at 16 UT), and -55 nT (June 28 at 8 UT). There was no significant increase in energetic particles during the June 18 CME above the SEP background due to an earlier SEP event from a west-limb CME on June 18 at 01:34 UT. The June 19 halo CME did not result in an SEP event. The June 21 CME was associated with an intense SEP event (1066 pfu in the >10 MeV GOES channel). The June 22 CME had high background from the June 21 SEP event and probably had a large SEP event (~200 pfu). The June 25 CME was also associated with a large SEP event (22 pfu). Thus the June 21 CME was a double-whammy event with a super geomagnetic storm and an intense SEP event.

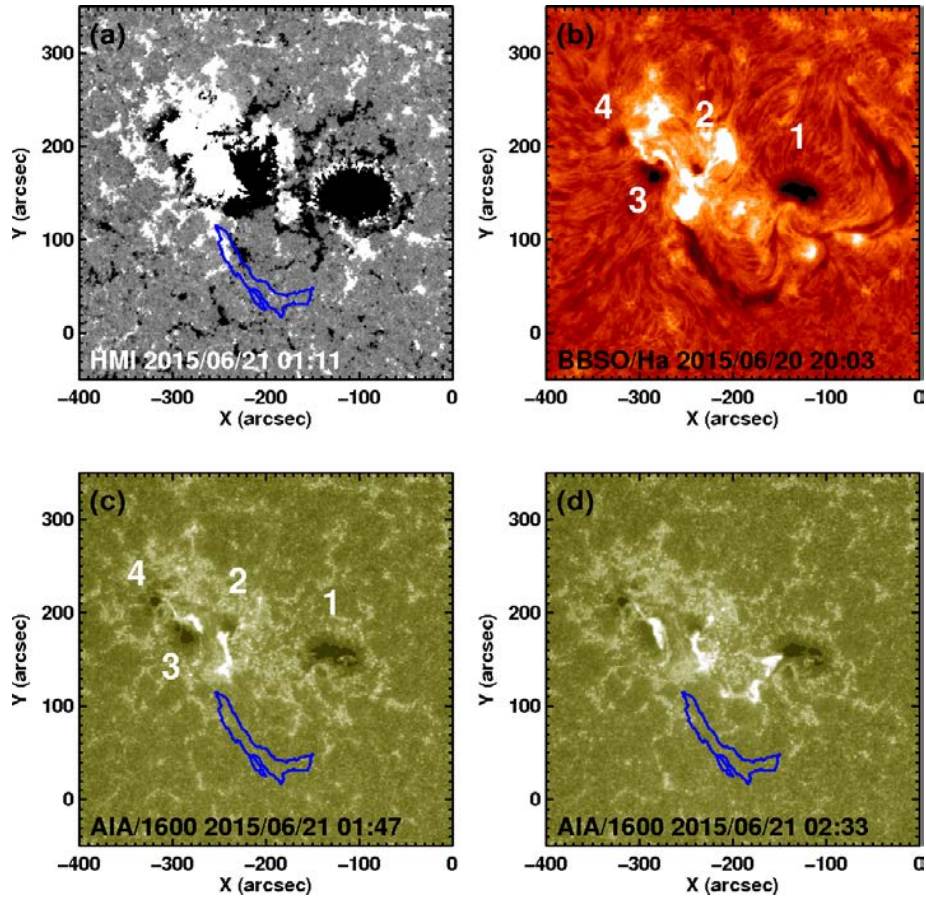

*Figure 1. Source region of the 2015 June 21 CME. (a) SDO/HMI line-of-sight magnetogram taken at 01:11 UT showing AR 12371 (white is positive and black is negative). (b) H-alpha image of the eruption region taken at 20:03 UT on June 20, a few hours before the flare onset. (c) An SDO/AIA 1600 Å image taken at 01:47 UT near the peak of the M2.0 flare showing the flare ribbons. (d) An SDO/AIA 1600 Å image taken near the peak of the M2.6 flare (02:33 UT). The quiescent part of the H-alpha filament is superposed on the images in other panels (blue contour). The leading (negative) part of the region had two sunspots at (-120", 160") and (-240", 185") and the trailing (positive) part also had two spots at (-315", 210") and (-285", 175"). We marked these spots as 1-4 going from west to east in (b) and (c).*

### 3.2 Solar Source of the June 21 CME

There were two M-class flares near the onset of the June 21 CME: (i) M2.0 flare starting at 01:02 UT with a peak at 01:42 UT, and (ii) M2.6 flare starting at 02:03 UT and peaking at 02:49 UT. Figure 1 shows an overview of the source region as observed in SDO/HMI magnetograms, Big Bear Solar Observatory's H-alpha image, and SDO/AIA 1600 Å images. The region was quite complex with an active region filament extending to become a quiescent filament shown in Fig. 1. The panels in Fig. 1 show that there were four sunspots in AR 12371, two with negative polarity (#1, #2) and two with positive polarity (#3, #4). The two negative spots were separated by a large distance (~120") with an intruding positive polarity patch between them. The positive polarity spots were close to each other aligned roughly in the north-south direction. The first flare involved both the positive polarity spots (#3, #4), but only the eastern negative polarity spot (#2). The second flare involved all the four spots (see the SDO/AIA 1600 Å images in Fig. 1). Because of this, the second flare covered a much larger area than the first one.

During the first flare, one can see brightening on either side of the neutral line between the positive spots (#3, #4) and the eastern negative spot (#2). The western negative spot (#1) was not involved in the flare. The 1600 Å ribbons are very short and separated by <20" because the opposite polarity patches were very close to each other. The ribbons started at the line joining the opposite polarity sunspots (#2, #3), but the eastern ribbon extended to the north while the western one extended to the south. The ribbons were also short (30 - 40"). During the second flare, there were also ribbons along the neutral line of the first flare, but the ribbons were more parallel, more separated, and extended to the south, In addition, the ribbon on the negative polarity side extended all the way to spot #1, taking a U-shape in going around the intruding positive magnetic patch between spots 1 and 2. The negative side of the ribbon was also located above the quiescent filament, which also erupted during the second flare.

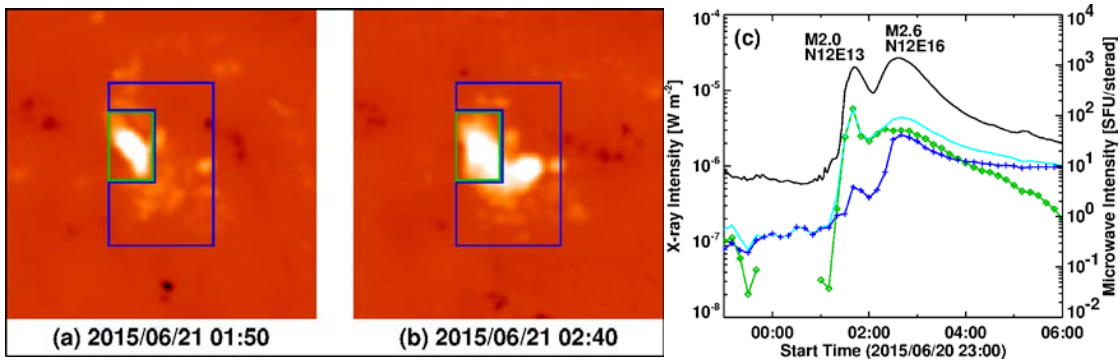

*Figure 2. The locations of the M2.0 (green box) and M2.6 (blue box) flares on 2015 June 21 as observed by the Nobeyama Radioheliograph at 17 GHz. The images correspond to the peak times of the flares. In (a) and (b) the sizes of the green and blue boxes are: 93".29×142".39 and 216".04×333".88. (c) GOES soft X-ray light curve along with the 17 GHz microwave intensity corresponding to the two flares. The green curve shows that at the location of the first flare, there was also emission from the second flare. The blue curve shows the variation in the blue box, excluding the emission from the green box. The microwave intensity in the entire region (including both flare boxes) is shown by the cyan line. The first flare was confined, while the second one was eruptive.*

The flares were imaged by the Nobeyama Radioheliograph at 17 and 34 GHz. We use the 17 GHz images made with an image cadence of 10 min here. The original data has much higher time resolution. The spatial resolution is about 10". Figure 2 shows 17 GHz images of the flare around the peak times of the flare. Consistent with what was observed in AIA 1600 Å images, the first flare is compact, while the second flare is extended. In order to see the time evolution of the microwave intensity in two areas, one corresponding the region of the first flare (green box) and the other an extended region including the second flare but excluding the area of the first flare (blue box). The intensity-time curve shows that the first flare is very impulsive, while the second flare is gradual. This is also seen in the blue curve, except that the emission corresponding to the first flare in the blue area is tiny, probably due to a remote-brightening associated with the first flare. The microwave intensities during the two flares was also reported in the online Solar Geophysical Data as 780 solar flux units (sfu) and 120 sfu for the first and second flares, respectively. The high microwave intensity is typical of confined flares as discussed below.

### 3.3 Metric Radio Bursts

Figure 3 shows a composite dynamic spectrum in the frequency range 30 to 2000 MHz that covers the June 21 two flares. The first flare is remarkably quiet in radio, except for the intense microwave emission shown in Fig. 2 and an isolated reverse-drift type III burst at ~01:35 UT, which is in the rise phase of the M2.0 flare. Reverse drift type III bursts are indicative of electron beams moving toward the Sun from the acceleration site. The starting frequency of the reverse-drift type III burst is close to 400 MHz, indicating the low height of the acceleration region. The higher-energy population of the accelerated electrons are responsible for the microwave emission shown in Fig. 2. The significance of the reverse drift burst is that there are no nonthermal electrons flowing into the IP space causing normal type III bursts. The key property of confined flares is that there is no upward flow of mass either as nonthermal electrons or as a CME (Gopalswamy et al. 2009b).

During the second flare, type III, type II, and type IV bursts were observed. According to the online Solar Geophysical Data, a type II burst was observed at the Palehua station of the Radio Solar Telescope Network (RSTN) in the frequency range 180 to 25 MHz. This is confirmed by the Culgoora observations shown in Fig. 3. The type II burst is quite faint compared to the type IV burst. The onset time of the type II burst was at 02:24 UT around 180 MHz and drifted down to at least 25 MHz by 02:47 UT. The type II burst was followed by a type IV burst with a lot of fine structures and lasted for many hours. The type II burst was faint but is an important indicator of CME-driven shock low in the corona. From the type II burst in Fig. 3, we can estimate a drift rate of 0.24 MHz/s between 180 and 80 MHz. This corresponds to a reference frequency of 130 MHz and we assume fundamental emission. The drift rate is consistent with the empirical relation between the emission frequency *f* and the drift rate (*df/dt*) reported by Gopalswamy et al. (2009c):

(1)    $|df/dt| = 4\times10^{-4} f^{1.27}$.

For $f$ = 130 MHz, the empirical relation gives $|df/dt|$ = 0.19 MHz/s, which is within the scatter of the observed data points employed in obtaining the empirical relation.

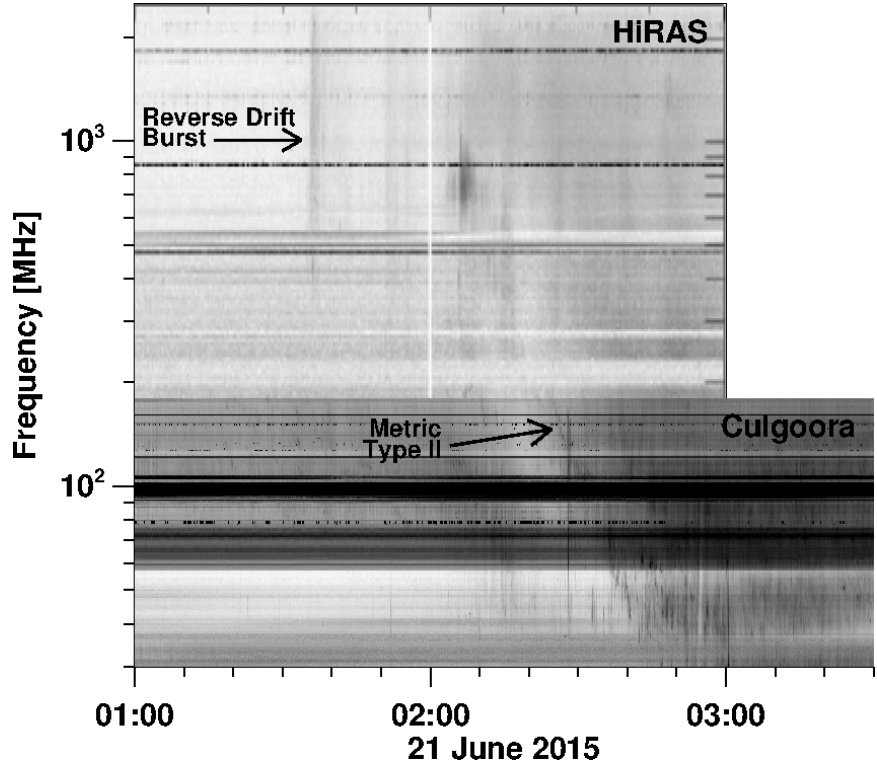

*Figure 3. A composite radio dynamic spectrum constructed using data from the Hiraiso Radio Spectrograph (HiRAS) and the Culgoora Radio Spectrograph. During the confined flare, only a reverse drift type III burst was observed. During the eruptive flare, type III, type II, and type IV bursts were observed. The type II burst started around 180 MHz and drifted down to at least 80 MHz. The dark continuum to the right of the type II burst (02:30 onwards) is the type IV burst with a lot of fine structures.*

The confined and eruptive characteristics of the two flares can also be inferred from the SDO/AIA images. Figure 4 shows two snapshot difference images at 193 Å. The image at 01:41 UT corresponds to the peak of the first flare. Although we saw some small-scale dimming at the northern and southern end of the flare site (probably due to loop expansion), there is no obvious disturbance spreading from the eruption site. On the other hand, the eruptive flare has extensive disturbance, which spread to a radius of ~0.5 Rs (solar radii) by 02:07 UT. The EUV disturbance is the early manifestation of the CME observed a little later in the LASCO field of view (FOV) at 02:36 UT.

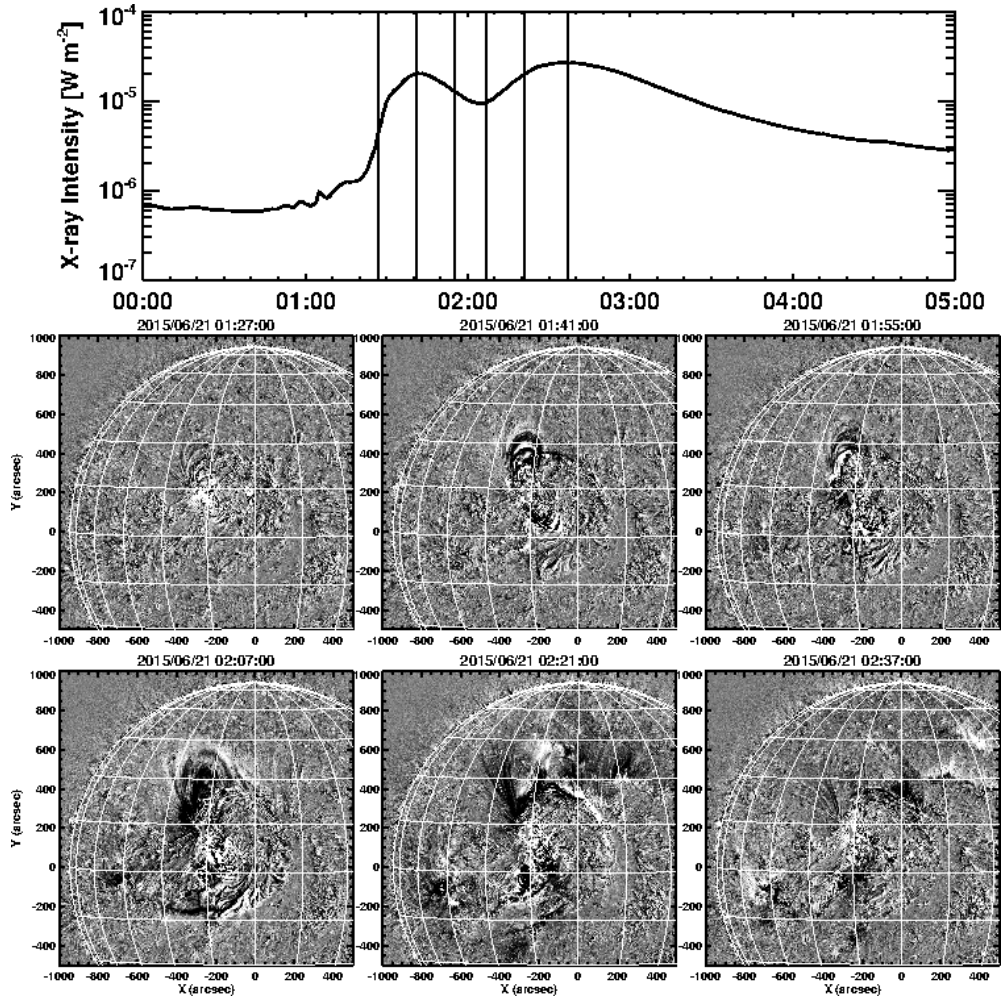

*Figure 4. SDO/AIA 193 Å difference images corresponding to the M2.0 and M2.6 flares. Each difference image is made by summing 5 long-exposure original SDO images taken over a period of 2 min and rebinning to a size of 1024x1024 pixels. The vertical lines on the GOES plot correspond to the times of SDO/AIA frames shown. The top three EUV images correspond to the duration of the M2.0 flare, while the bottom three correspond to the M2.6 flare. The dimming in the top EUV images do not show any spreading, indicating no eruption. The disturbance during the M2.6 flare clearly spreads and was associated with the white-light CME, which first appeared in the LASCO FOV at 2:36 UT (right). We confirmed the confined and eruptive nature by playing movies of EUV images. The disturbance at 02:07 UT has a radius of ~450", indicating the CME leading edge is at a height of 1.5 Rs from the Sun center.*

### 3.4 CME Kinematics

The white-light CME appeared at a height of 3.53 Rs at 02:36 UT above the northwest quadrant, but quickly became a halo CME in the next frame taken at 03:48 UT with its leading edge advancing to 4.86 Rs. Clearly this was a fast CME because the sky-plane speed is more than 1200 km s$^{-1}$ between these two points. The leading edge is actually the CME-driven shock seen as a diffuse feature (see Fig. 5) surrounding the bright CME flux rope. The shock structure is consistent with the high speed of the leading edge. The height-time plots available at the SOHO/LASCO CME catalog indicates an average speed of 1366 km s$^{-1}$ in the LASCO FOV. A second-order fit to the height-time data points suggests that the CME continued to gain speed in the LASCO FOV with an average sky-plane acceleration of ~21.2 m s$^{-2}$ and reaching a speed of 1477 km s$^{-1}$ by the time the leading edge left the LASCO FOV.

Even though there were no STEREO observations available for this CME, we were able to fit a flux rope and a shock to the SOHO/LASCO white-light features using the Graduated Cylindrical Shell (GCS) model (Thernisien 2011). The fitting routine RTCLOUDWIDGET.PRO has been implemented in *Solarsoft*. A spheroid shock model was added to this code (Olmedo et al. 2013). The shock model has six free parameters, four of which are the same as in the flux rope model: propagation direction (latitude, longitude), leading-edge distance, width, and tilt angle. The remaining two parameters are the major and minor axes of the spheroid. Details can be found in recent works that applied the shock-rope model combination to track the leading edges of the shock and flux rope (Hess and Zhang, 2014; Mäkelä et al. 2015; Xie et al. 2017).

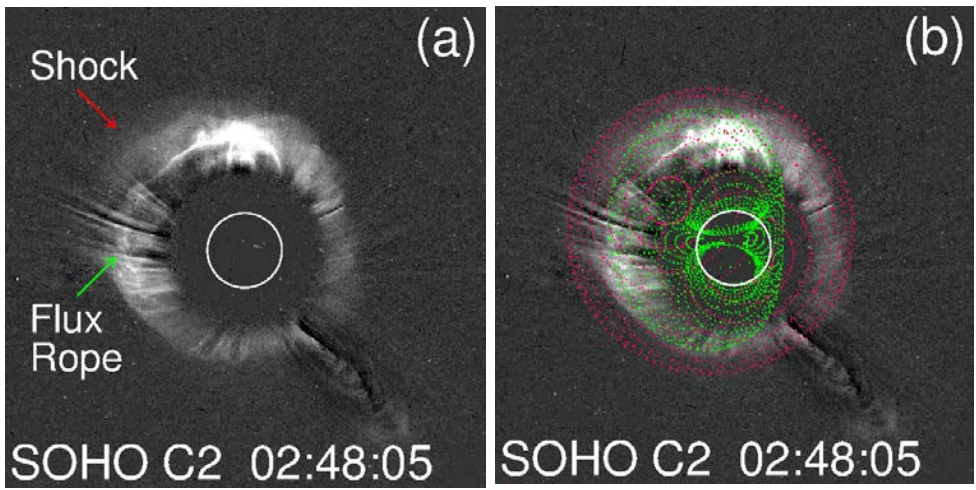

*Figure 5. (a) The white-light difference image at 02:48 UT with the flux rope and the shock feature marked. (b) GCS model fit to the white-light features (green: flux rope; red: shock). The face-on and edge-on half widths of the flux rope are 92º and 37º, respectively with an average half width of the CME is 55º. The flux rope field has a left-handed helicity as expected for the northern hemisphere, consistent with the filament orientation. The tilt angle is -85º, indicating that the axis is to the left of the vertical in the northern hemisphere. The propagation direction of the flux rope is N11E25 and the kappa parameter is 0.75 (see text).*

The results of the flux rope and shock fitting are shown in Fig. 5. The flux rope axis has roughly a north-south orientation (tilt angle is -85º), consistent with the vertical filament and the main neutral line involved in the eruption. The fitted direction is N11E25 shifted to the east from the flare direction. The fitted height of the shock at 02:48 UT is 6.18 Rs compared to the sky-plane height of 4.86 Rs. By fitting flux ropes and shocks to successive frames in white light and EUV, we obtained the height time history of the shock.

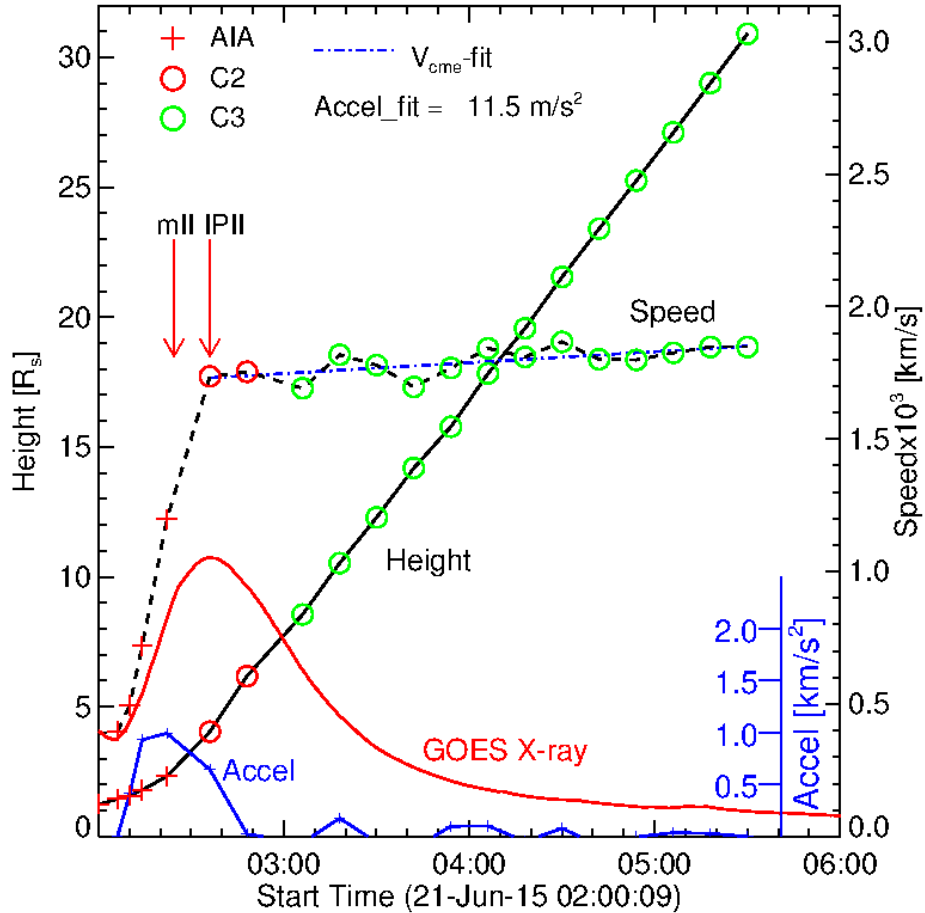

*Figure 6. Height-time, speed-time, and acceleration-time plots of the shock obtained from the GCS fit. (a) Measurements from SDO/AIA are denoted by plus symbols. Measurements from LASCO/C2 and LASCO/C3 are shown by red and green dots, respectively. A linear fit to the LASCO data points alone yields an average speed of 1784 km s$^{-1}$. Shock speed and acceleration obtained from successive data points. The speed rapidly increases in the first half hour, and then slowly afterwards. The slowly increasing part has an average acceleration of 11.5 m s$^{-2}$ with a final speed of 1842 km s$^{-1}$ in the LASCO FOV. The acceleration peaked close to ~1 km s$^{-2}$. The 1-8 Å GOES light curve is shown for reference. The start times of metric (mII) and interplanetary (IPII) bursts are marked.*

Figure 6 shows the time variation of the shock height obtained from the shock fit to the EUV and LASCO images. The speed and acceleration are obtained from successive data points. The acceleration obtained from the shock speed also applies to the flux rope because they are tightly coupled near the Sun. A linear fit to the LASCO data points gives an average speed of 1784 km

s$^{-1}$ within the LASCO FOV. There is clear initial acceleration between the SDO and LASCO FOVs. By fitting a second order polynomial to the SDO and LASCO/C2 data points in Fig. 6, we get an average acceleration of 784 m s$^{-2}$. The speed evolution is shown in Fig. 6 using successive height-time data points. The speed increases rapidly until about 02:35 UT and then increases slowly. The acceleration derived from successive speed points peaks at 02:20 UT with a peak value of 960 m s$^{-2}$, not too different from the quadratic fit to the early heights in Fig. 6. The CME speed evolution in Fig. 6 is somewhat similar to the soft X-ray increase as was pointed out by Zhang et al. (2001). The slower increase in speed corresponds to an average acceleration of 11.5 m s$^{-2}$. By the time the CME left the LASCO FOV, the shock speed was 1894 km s$^{-1}$.

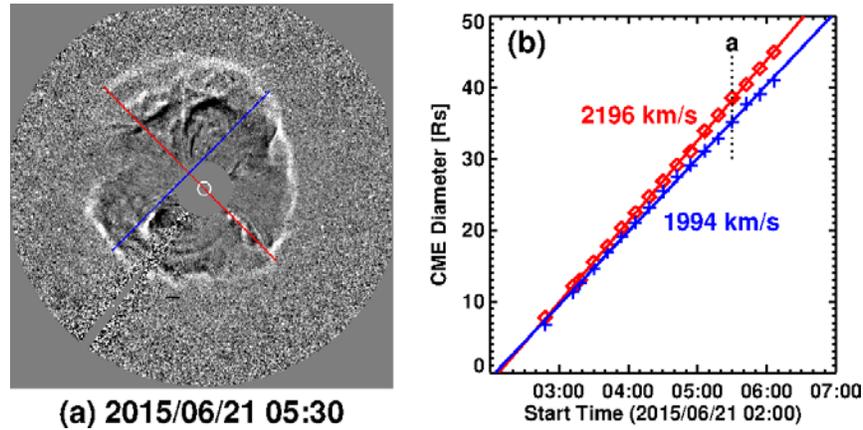

*Figure 7. (a) A LASCO C3 image at 05:30 UT with the CME diameters drawn at two locations, connecting position angles 45º and 225º for the red line and 135º and 315º for the blue line. The time variation of this diameter gives the expansion speed of the CME. (b) The expansion speeds measured at the two pairs of position angles. The average of the two expansion speeds is 2095 km s$^{-1}$. The time corresponding to the image in (a) is marked by the vertical dotted line.*

In addition to the GCS fit, we have other means to confirm the high speed of the CME. Since the CME was a full halo, one can readily obtain the 3-D speed using a cone model. Such a speed is already reported in the SOHO/LASCO Halo CME catalog (https://cdaw.gsfc.nasa.gov/CME_list/halo/halo.html, Gopalswamy et al., 2010b). The listed value of 1740 km s$^{-1}$ is not too different from the average speed 1784 km s$^{-1}$ obtained from the GCS fit. For full halo CMEs, it is straight forward to measure the expansion speed ($V_{exp}$) and then derive the radial speed ($V_{rad}$) from it (Schwenn et al. 2005; Gopalswamy et al. 2009d). Figure 7 illustrates the expansion speed measurements from the increase in CME diameter at two position angles. The expansion speeds of the CME at position angles 45º and 135º are 2196 and 1994 km s$^{-1}$, respectively, giving an average value of 2095 km s$^{-1}$. If we apply the empirical formula (Schwenn et al. 2005),

(2)    $V_{rad} = 0.88\ V_{exp}$,

we get $V_{rad}$ = 1844 km s$^{-1}$, which is also consistent with the speed in Fig. 6. To account for the width dependence of the expansion speed, Gopalswamy et al. (2009d) used the full ice cream cone model with a cone half width $w$ to obtain the relation,

(3)     $V_{rad} = f(w) V_{exp}$,

where $f(w) = ½(1+\cot w)$. We know $w$ from the flux rope fit (55º), so $f(w) = 0.85$, giving $V_{rad}$ = 1781 km s$^{-1}$ in good agreement with the linear speed in Fig. 6. Finally, Gopalswamy et al. (2015a) found an empirical relation between sky-plane ($V_{sky}$) and 3-dimensional ($V_D$) speed of CMEs:

(4)     $V_{3D} = 1.1 V_{sky} + 156$.

Substituting $V_{sky}$ = 1366 km s$^{-1}$, we get $V_{3D}$ = 1659 km s$^{-1}$, only slightly below the cone model speed, but within the typical measurement errors (~10%). Piersanti et al. (2017) assumed (based on the CORIMP automatic CME catalog) that the CME slowed down to ~300 km s$^{-1}$ in the outer part of the LASCO FOV (beyond our measurements). The above analyses show that there is no evidence for such a low speed. The low speed obtained by these authors seems to be an artifact of the automatic CME measurements. We will show later that the slowdown in the IP medium is gradual, but the CME speed always remained much higher than the slow solar wind speed. We will further investigate the speed evolution of the shock in the IP medium using the drift rate of the well-observed IP type II radio burst by the Wind/WAVES instrument.

### 3.5 Interplanetary Type II Burst and Shock-speed Evolution

The strongest evidence for an energetic CME in the IP medium is the existence of an intense type II burst in the decameter-hectometric (DH) to km wavelengths. In addition, there was a metric type II component reported in section 3.2. Type II bursts occurring at such wide frequency range (m to km) are caused by strong shocks driven by the most energetic of CMEs (Gopalswamy et al. 2005a; 2010c; 2012; Cremades et al. 2015). CMEs associated with such m-km type II bursts are also known to cause large SEP events (Gopalswamy 2006; Gopalswamy et al. 2008). Figure 8 shows the Wind/WAVES dynamic spectrum over 48 hours from 0 UT on June 21 to the end of the day on June 22. During this interval, there were three shocks (S1, S2, S3) detected by Wind/WAVES. S1 and S2 were due to previous CMEs, while S3 on June 22 at 18:01 UT is due to the CME in question.

The dynamic spectrum shows that the event was associated with a couple of DH type III bursts, one at 02:02 UT close to the onset of the M2.6 flare and the second type III at 2:27 UT, just before the onset of the Type II burst at 02:36 UT. The starting frequency of the type II burst was in the frequency range 5-6 MHz. The instantaneous bandwidth of the type II burst was very large suggesting that a large section of the shock was producing type II bursts. Only a few events in each solar cycle typically have such broadband signature (Gopalswamy et al. 2003). The broadband nature also indicates that the fundamental and harmonic components of the type II burst merged during most of the time except for a few short intervals. Even while arriving at the Wind spacecraft, the shock was emitting both fundamental (~35 kHz) and harmonic (~70 kHz) components as can be seen in Fig. 8. Type II bursts near the observing spacecraft have been reported before from Wind and STEREO observations (Bale et al. 1999; Reiner et al. 2007; Liu et al. 2013; Schmidt and Cairns 2015; Gopalswamy et al. 2016).

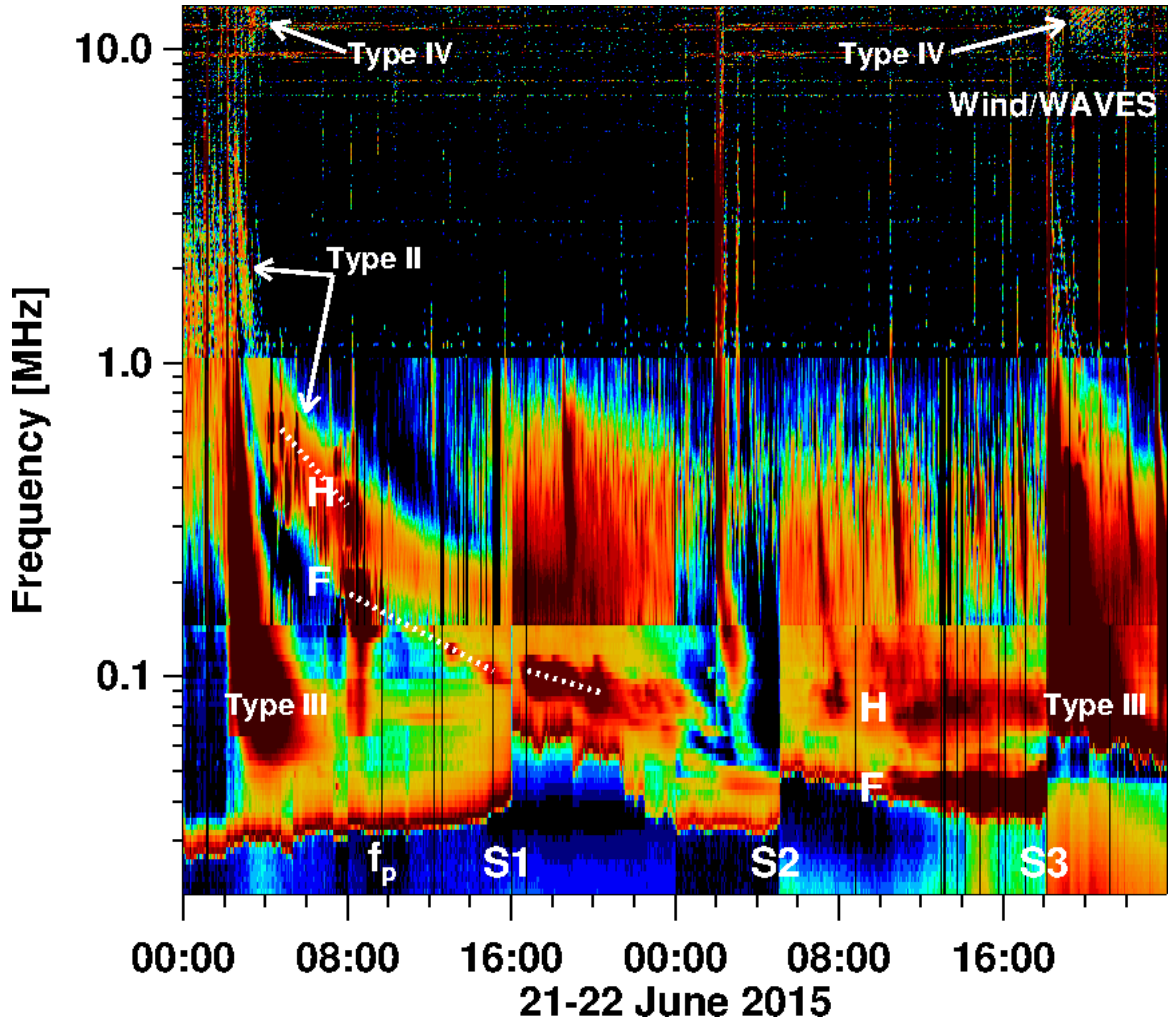

*Figure 8. A composite dynamic spectrum from Wind/WAVES showing the IP type II burst from DH to km wavelengths. The spectrum has three parts: RAD2 (above 1 MHz), RAD1 (0.15 MHz to 1 MHz), and TNR (below 0.15 MHz). Near the onset of the type II burst, there was a DH type III burst. F and H denote the fundamental and harmonic components of the type II burst. There are two type IV bursts (marked by arrows) near the upper edge of the dynamic spectrum. The first one (upper left) is associated with the June 21 CME. The second type IV burst to the upper right is from the June 22 CME. The dotted lines mark sections of the F and H components where drift rate measurements were made. S1, S2, and S3 are three IP shocks observed by the WAVES/TNR instrument as a jump in the plasma line (thermal radio noise detected at the local plasma frequency $f_p$). S3 arrived at 18:01 UT on June 22 and is due to the CME in question. The type III*

*burst immediately after S3 is due to another halo CME from the same region that occurred on June 22 at 18:36 UT.*

In addition to the type III and type II bursts, the extension of the metric type IV burst was observed down to 10 MHz in association with the June 21 CME. The homologous CME on June 22 also was associated with a DH type IV burst with a slightly longer duration as indicated in Fig. 8. DH type IV bursts are rare events associated with very energetic CMEs with an average sky-plane speed of ~1500 km s$^{-1}$ and most of them (78%) are halo CMEs (Gopalswamy 2011). The June 21 CME properties are consistent with these properties. Clearly, the presence of type IV burst during the 2015 June 21 CME is another strong evidence of an energetic CME.

The drift rate of type II bursts carries important information on the density of the ambient medium because the emission takes place at the fundamental (F) and second harmonic (H) of the local plasma frequency. We computed the drift of the June 21-22 type II burst at several intervals as listed in Table 2. If a burst drifts from frequency *f1* to *f2* over a time interval *dt*, then the drift rate d*f*/d*t* is given by (*f2-f1*)/*dt*. This quantity is negative for type II bursts because the drift is from higher to lower frequencies (*f1* > *f2*). For the fundamental component the frequency *f* of radio emission is at the local plasma frequency

(5)     $fp$ (MHz) = $9.0 \times 10^{-3} n^{1/2}$,

where *n* is the local plasma electron density in cm$^{-3}$. Therefore d*f*/d*t* is the same as d*fp*/d*t*. One can convert the measured d*f*/d*t* into the shock speed as follows:

(6)     d*f*/d*t* = d*fp*/d*t* = $9.0 \times 10^{-3}$ (½$n^{-1/2}$)(d*n*/d*t*) = ($fp$/2)(1/*n*)(d*n*/d*r*)(d*r*/d*t*),

where we have used the definition in Eq. (5).

The factor d*r*/d*t* is the shock speed *V* and |(1/*n*)(d*n*/d*r*)| is the inverse of the density scale height *L* of the ambient medium. Thus, the drift rate of the type II burst contains information on the shock as well as on the ambient medium. Knowing the density scale height, we can get the shock speed by rearranging Eq. (6) as

(7)     *V* = 2*L*(dln*f*/d*t*).

Note that we can measure the normalized drift rate (dln*f*/d*t*) from the dynamic spectrum. All we need is the density scale height *L* to get the shock speed. Assuming a density variation of the form

(8)     $n = n_0 r^{-\alpha}$,

where $n_0$ is the density at some reference distance we get the scale height as,

(9)     *L* = *r*/α.

Therefore, Eq. (7) becomes

(10) $V = (2r/\alpha)(\mathrm{d}\ln f/\mathrm{d}t)$.

It is well known that $\alpha = 2$ in the IP medium, valid typically beyond 10 Rs from the Sun. In this case the shock speed is simply the product of the heliocentric distance and the normalized drift rate:

(11)  $V = r(\mathrm{d}\ln f/\mathrm{d}t)$.

Near the Sun, $\alpha$ is large: ~6 or larger in the inner corona. In the outer corona, where DH type II bursts start, $\alpha$ ~4 that slowly drops to 2 in the IP medium. At distances where white-light shock measurements are available, we can estimate $\alpha$ using $V$ and $\mathrm{d}\ln f/\mathrm{d}t$ from observations. At distances beyond the coronagraph FOV, we can calibrate the densities from the emission frequencies assuming $r^{-2}$ density fall off.

*Table 2. Speed evolution of the CME driven shock from the Sun to 1 AU*

| Time Range (UT) | $f1, f2$ (MHz) | $n$ (cm$^{-3}$) | $(1/f)(\mathrm{d}f/\mathrm{d}t)$ (s$^{-1}$) | $\alpha$ | $r$ (Rs) | $V$ (km s$^{-1}$) |
|---|---|---|---|---|---|---|
| 02:34:08 to 03:06:40 | 3.2, 1.0 | 54444 | $-5.36\times10^{-4}$ | 2.7[c] | 6.4 | 1750[d] |
| 03:13:00 to 05:04:15 | 0.96, 0.37 | 5475 | $-1.34\times10^{-4}$ | 1.9[c] | 18.2 | 1842[d] |
| 04:47:07 to 07:57:54 | 0.611, 0.350[b] | 714 | $-4.73\times10^{-5}$ | 2.0 | 50.4 | 1658[e] |
| 08:08:05 to 15:06:11 | 0.18, 0.10 | 254 | $-2.19\times10^{-5}$ | 2.0 | 84.7 | 1289[e] |
| 16:49:00 to 20:35:00 | 0.104, 0.088 | 114 | $-1.24\times10^{-5}$ | 2.0 | 126.0 | 1089[e] |
| 06/22/2015 18:01[a] | ---- | 19.8 | ---- | ---- | 214.0 | 776[f] |

*[a]Time of the shock at the Wind spacecraft. [b]Harmonic components. [c]Assuming shock speed derived from radio and white light to be the same. [d]From Wind/WAVES. [e]Shock speed derived from type II drift rate and density scale height. [f]In-situ shock speed at Wind.*

Table 2 shows the drift rate measurements at several time intervals of the radio dynamic spectrum and the corresponding distances from the Sun. It is assumed that the radio emission comes from the nose region of the shock. However, at 1-AU, there are emissions extending to higher frequencies from the local fundamental-harmonic components suggesting emission from the nose and flanks. For the first two distances, the shock speeds are available from the shock fit to the white light observations. In these two cases, we obtain the exponent $α$ as 2.7 at 6.4 Rs and 1.9 at 18.2 Rs. These numbers are consistent with the expected values noted above. For the segments of the dynamic spectra starting at 8:08 and 16:58 UT, we derive the densities from the emission frequency ($= fp$) and use the relation (8), by taking $n_0$ to be the density (5475 cm$^{-3}$) at the reference distance of 18.2 Rs. Finally, we have also given the shock speed at the Wind spacecraft as determined from various methods using the Wind data and made available at the web site, https://www.cfa.harvard.edu/shocks/wi_data/00693/wi_00693.html.

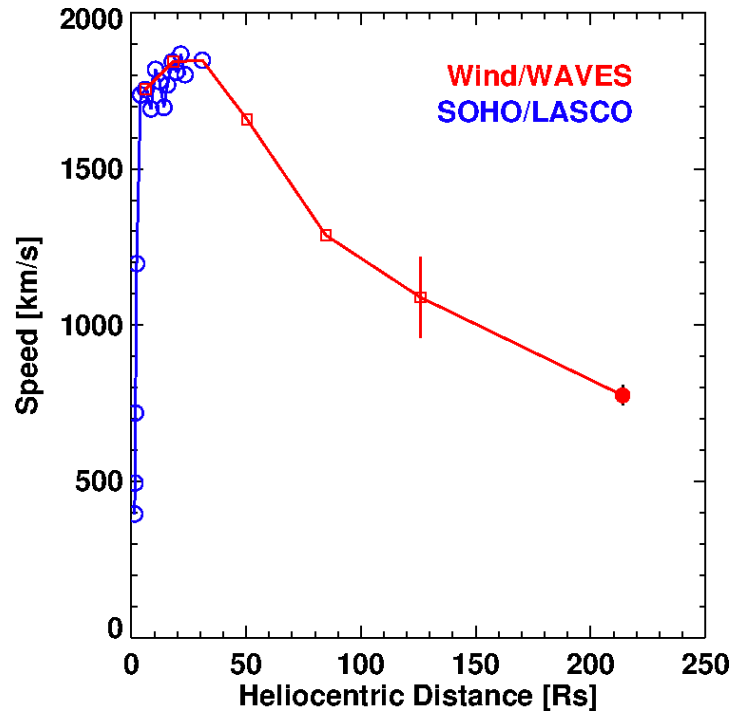

*Figure 9. Radial evolution of the shock speed from the Sun to Earth. Data from the GCS shock fit (EUV + white light observations shown by the blue circles), from Wind/WAVES radio dynamic spectrum (red squares), and the in-situ shock speed from Wind data (solid circle). The error bar on the speed at r = 126 Rs was because of the uncertainty in defining the slope: a 1% shift in up or down of the end frequencies resulted in a speed shift of 130 km s$^{-1}$. The shock speed from Wind (775.8 km s$^{-1}$) is an average of results from eight different methods ranging from 698 km s$^{-1}$ to 793 km s$^{-1}$. The error bar is the standard deviation of the eight measurements. It must be noted that Wind might have sampled the shock flank, in which case the nose speed is likely to be higher.*

The radial evolution of the shock speed is shown graphically in Fig. 9, which combines the speed evolution in the corona obtained from white-light observations (see Fig. 6) and that in the IP medium from Table 1. We see that the shock speed rapidly increased in the inner corona, slowly increased in the outer corona, and then started decreasing in the inner heliosphere where the drag force is supposed to dominate. Interestingly, the speed decreased only slightly in the second half of the Sun-Earth distance. This is consistent with the acceleration cessation distance inferred in Gopalswamy et al. (2001a) from Helios observations and in Poomvises et al. (2012) from STEREO Heliospheric Imager observations. The slowdown of IP shocks is known for a long time (see e.g., Gosling et al. 1975; Woo et al. 1985), but we are now able to characterize it better using heliospheric information provided by IP radio data and Heliospheric Imager data (see e.g., Reiner et al. 2007; Liu et al. 2013; Hess and Zhang 2017).

### 3.6 In situ Observations and Geomagnetic Storm

Figure 10 shows the solar wind conditions over a three-day period around the time of arrival of shock S3, including the shocks S2 and S4 (see Table 1 and Liu et al. 2015). The shock S3 was followed by two small ICME intervals (ICME1, ICME2) that lasted only for about 4 hours each. The shock standoff distance was about 7.5 h. There was an intervening high-temperature structure between ICME1 and ICME2, well above the expected solar wind temperature shown for comparison (Lopez and Freeman 1986). ICME1 and ICME2 intervals also do not have the temperature below the expected solar wind temperature, but the plasma beta was low. The densities in the two intervals were low, but that during ICME2 was unusually low (see the density plot in Fig. 10). The low density was also observed in the Advanced Composition Explorer (ACE) data (see Marubashi et al. 2017). There was an extended interval with low proton temperature that looks like ICME material beyond 16:00 UT on June 23 until the arrival of shock S4 at ~13:00 UT on June 24. Both the temperature and plasma beta of this interval are low in contrast to the intervals ICME1 and ICME2 that have low plasma beta but not the low temperature. This suggests that the extended ejecta material may be of different origin and was separated from the interval ICME2 by another hot structure that lasted for ~4 h. The magnetic field strength of the low-temperature interval was in the range 5-10 nT, barely above the solar wind value. Liu et al (2015) treated this interval combined with the intervals of ICME1, ICME2, and the hot structures as a single ICME interval lasting for ~36 h.

There was also no rotation in the By and Bz components during ICME1 and ICME2: Bz points to the south and By points to the east throughout the intervals. Therefore, fitting a typical cylindrical magnetic cloud model (e.g., Lepping et al. 1991) is not possible for these intervals. However, the speed profile declines from the front boundary of ICME1 to the back boundary of ICME2 indicating expansion of a single flux rope (if the hot structure is ignored in Fig. 10). Liu et al. (2015) were able to fit flux ropes using the Grad-Shafranov technique to the intervals ICME1 and ICME2, both of which had left-handed (LH) chirality. Liu et al. (2015) suggest that a single flux rope was split into two by the intervening hot structure. The LH chirality inferred by Liu et al. (2015) agrees with the helicity sign of the GCS flux rope derived in section 3.4.

Marubashi et al. (2017) presented ACE data, which indicate that the hot intervening structure was very weak, appearing as a fluctuation. Therefore, Marubashi et al. (2017) fitted a single flux rope to the interval from the beginning of ICME1 (01:30 UT on June 23) to the end of ICME2 (11:30 UT on June 23). They obtained two flux-rope solutions: one with RH chirality and the other with LH chirality. The RH solution has the right inclination but wrong chirality. The LH solution has the right chirality, but wrong inclination.

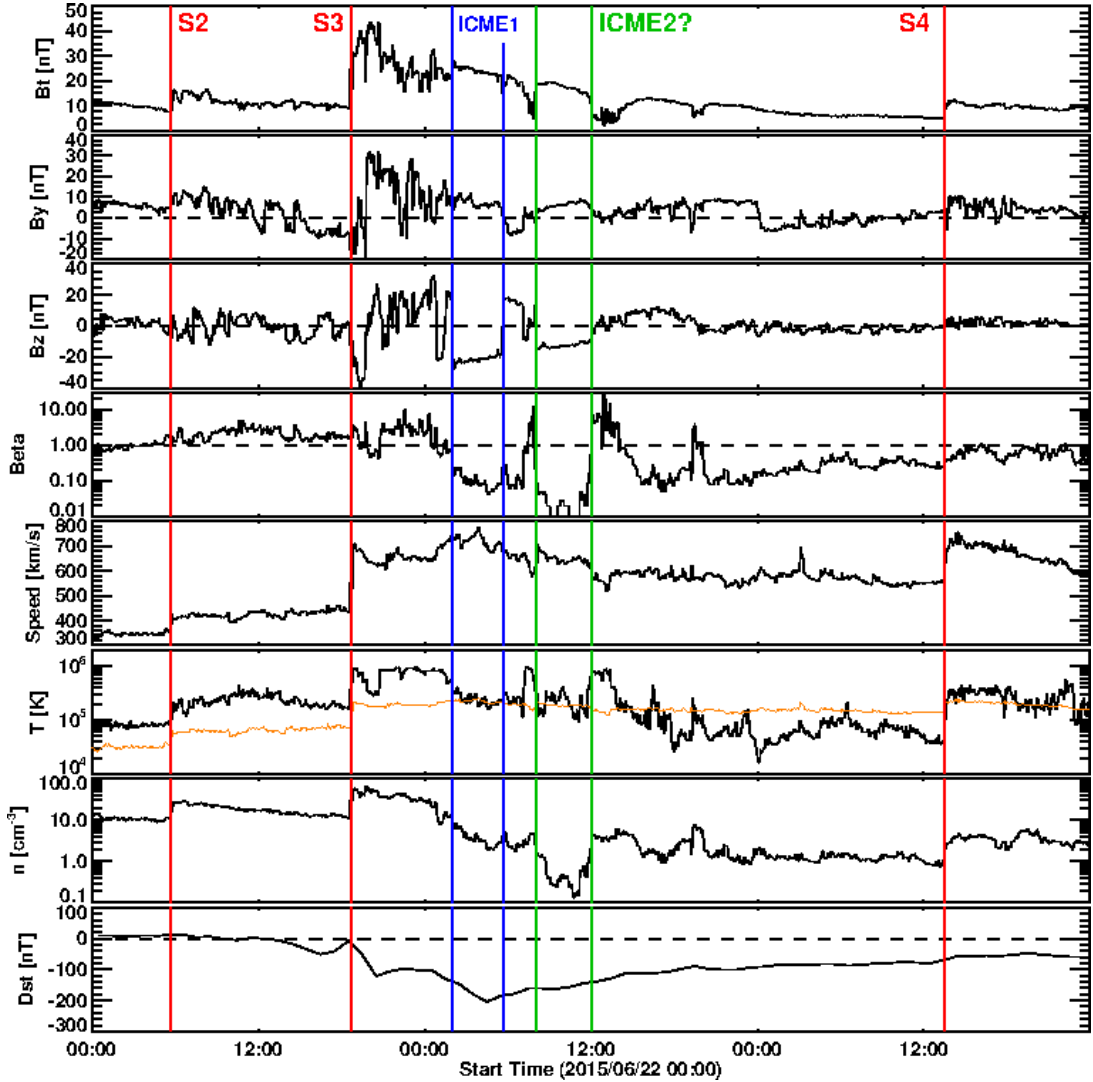

*Figure 10. Solar wind conditions around the time of arrival of the shock (S3) associated with the 2015 June 21 CME. Plotted from the top to bottom: total magnetic field strength, components By and Bz, plasma beta, solar wind speed, solar wind temperature (T), solar wind density (n) and the Dst index. In the temperature plot, we have also shown the solar wind expected temperature in orange. The vertical blue lines indicate the duration of an ICME (ICME1). The green solid lines indicate a possible second ICME interval (ICME2), which may be part of the same ICME split into two by a hot intervening structure. The ICME speed ranges between 740 km s$^{-1}$ at the leading edge of ICME1 and 620 km s$^{-1}$ at the trailing edge of ICME2, giving an average speed of*

*680 km s$^{-1}$. Shocks S2 and S4 are also shown for context. The low-temperature interval from about 16 UT on June 23 to the arrival of S4 may not be related to the preceding ICME.*

In summary, we suggest that the interval from the beginning of ICME1 to the end of ICME2 is a single flux rope. This is in agreement with Liu et al. (2015) and Marubashi et al. (2017). However, we think the low-temperature ejecta beyond the end of ICME2 does not belong to the 1-AU counterpart of the June 21 CME. It is likely due to eruptions that occurred in the vicinity of AR 12371 in the interval between the June 21 and June 22 halo CMEs. Thus the difference between previous work and ours is in the overall ICME interval: Liu et al. (2015) think that the overall ICME duration of 36 h is consistent with head-on collision. Our estimate of the total duration of the flux rope is only about 10 h, which is much smaller than usual (the average duration of magnetic clouds is ~20.5 h, see Gopalswamy 2008). The small size seems to be due to the deflection of the flux rope away from the Sun-Earth line by a large equatorial coronal hole (see section 4.1 below). Based on the duration alone, a head-on collision scenario is plausible, but the in-situ properties are not compatible with a coherent flux rope obtained from the GCS fit.

Even though there have been multiple halo CMEs and their shocks that occurred around the June 21 CMEs, they were generally well isolated. The only significant effect of these multiple events is that the upstream density of S3 was higher because S3 was propagating into the extended sheath of the preceding shock S2. This is obvious from the emission frequencies of the fundamental and harmonic bands at the Wind spacecraft: ~35 kHz and ~70 kHz, which are higher than in normal solar wind at the Wind spacecraft (see Fig. 8). The average plasma density in the upstream of S3 is ~15 cm$^{-3}$, corresponding to a plasma frequency of ~ 34.9 kHz. This is consistent with the lower edge of the fundamental band of the type II burst shown in Fig. 8.

As for the geomagnetic activity associated with the June 21 CME, both the shock sheath and the ICME were geoeffective. The sheath had initially a strong southward $Bz$ component (~-40 nT) that caused the initial dip in Dst to major storm level (-121 nT). There was a smaller $Bz$ south excursion (-25 nT) at the back end of the sheath, which slightly resisted the recovery of the Dst index. The ICME field was all southward ($Bz$ ~ -25 nT), which caused the main dip in Dst to -204 nT, making it the second largest storm in cycle 24. We estimated the storm level from the empirical relation (Gopalswamy 2010),

(12)    $Dst$ = -0.01$VBz$ – 32 nT,

where $Bz$ is the magnitude of the southward component. For $V$ = 740 km s$^{-1}$ and $Bz$ = 25 nT, we get $Dst$ = -217 nT, closely matching the observed $Dst$ value. In this sense, there is nothing unusual about this storm; it is a direct consequence of the fast ICME structure with a moderate field strength. The CMEs were relatively well separated to have a significant change on the depth and duration of the storms. The June 22 and 25 CMEs occurred during the recovery phase of the superstorm, which likely prolonged the recovery phase of the superstorm.

## 4. Discussion

In this section, we explain some peculiarities of the CME/shock event and factors that might have affected the CME/shock evolution. In particular, we consider CME deflection by a large coronal hole close the CME source location and two other smaller coronal holes. Since we have multiple CME/shock speed measurements in the inner heliosphere, we estimate the IP deceleration and compare it with the empirical relation reported elsewhere. The low starting frequency of the IP type II burst seems to be contradictory to the presence of metric type II burst at very high frequencies (180 MHz). We provide an explanation that is consistent with the same CME-driven shock causing both radio emissions. The explanation is also consistent with the observed kinematics of the CME. Finally, we show that the coronal flux rope constructed from the FRED technique (Gopalswamy et al. 2017b,c) is consistent with the ICME properties that led to the second largest geomagnetic storm in cycle 24 (Gopalswamy et al., 2009e).

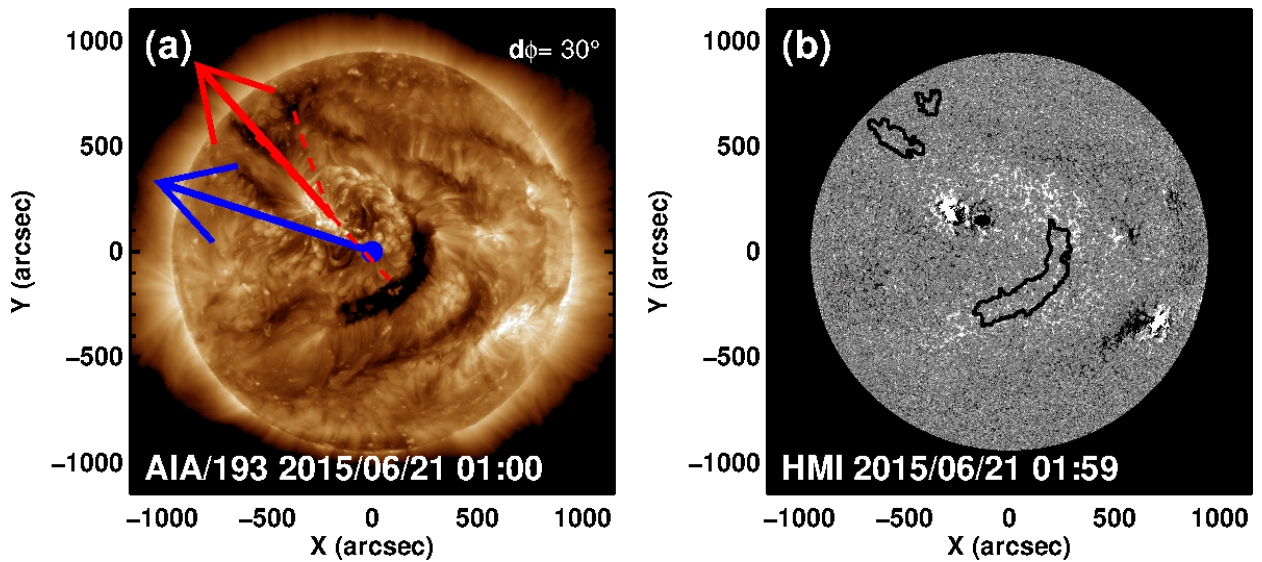

*Figure 11. CME deflection by coronal holes. (a) SDO/AIA image taken just before the eruption on 2015 June 21. The CHIP was computed for the three coronal holes connected to the eruption region by the red dashed lines. The red arrow indicates the direction along which the resultant coronal hole influence acts. The blue arrow is along the central position angle of the CME. The positing angle separation (dΦ) between CME central position angle and the position angle of coronal-hole influence is 30º. (b) SDO/HMI magnetogram taken at 01:59 UT with the contours of the three coronal holes (S07W06, N37E47, N50E39) overlaid. The magnetogram indicates that the polarity is positive inside all coronal holes.*

### 4.1 CME deflection by the Equatorial Coronal Hole

Now we revisit the short duration (~10 h) of the ICME, which is unusually small for a source location close to the disk center (N12E13). It is known that some CMEs originating near the disk center end up not detected as an ICME at 1 AU, even though the associated shock is detected. Such shocks are known as "driverless" shocks (Gopalswamy et al. 2009e). It was found that in these events, a coronal hole is present near the eruption region such that it deflects the CME sufficiently away from the Sun-Earth line and hence is not detected at 1 AU. A milder version of this deflection is that only a small section of the ICME is detected at 1 AU. Some examples of such short-duration ICMEs were reported earlier (Gopalswamy et al. 2012). We suggest that such a deflection is likely to have occurred in the present event. The evidence for such a deflection is presented in Fig. 11, where we have identified the coronal holes in SDO/AIA 193 Å image and shown them on a SDO/HMI magnetogram. There was a crescent shaped dark coronal hole near the disk center (S07W06) and two smaller coronal holes (N36E46, N50E39) to the northeast of the eruption region. We computed the coronal hole influence parameter (CHIP) as the product of the coronal hole area, the average photospheric magnetic field strength within the coronal hole, and the inverse square of the distance between the coronal hole and the eruption region.

The CHIP for the three coronal holes in Fig. 11 is listed in Table 3. For each coronal hole, the area, average field strength (<*B*>), and the distance of the eruption region to the coronal hole are listed. The computed magnitudes of X and Y components (*F*x, *F*y) as well as the resultant (*F*) of CHIP are also given for each coronal hole. The last column in Table 3 gives the position angle (*F*pa) along which the coronal hole influence acts. The coronal hole near the disk center (S07W06) was the most influential pushing the CME in the north-west direction, at a position angle of 41° with a CHIP of 12.0 G. The other two coronal holes had a smaller influence with CHIPs of 1.2 G (N37E47) and 0.0063 G (N50E39). The resultant CHIP is 11.0 G acting along position angle 42°. The position angle along which the CHIP acts is close to the central position angle of the CME (72°). Such small difference in position angles and a large CHIP value are characteristic of coronal holes that significantly deflect CMEs as was shown by Gopalswamy et al. (2009e). The main effect of the deflection of the CME in the north-east direction is that the observing spacecraft passes through the western flank of the ICME. Another effect is to reduce the earthward speed of the CME and hence delay the arrival time of the shock.

*Table 3 CHIP factor contribution from the three nearby coronal holes*

| Location | Area (km$^2$) | <B> (G) | Dist. (km) | Fx (G) | Fy (G) | F (G) | Fpa (deg.) |
|---|---|---|---|---|---|---|---|
| S07W06 | 3.9×10$^{10}$ | 5.7 | 3.2×10$^5$ | -8.0 | 8.9 | 12.0 | 41 |
| N37E47 | 2.1×10$^{10}$ | 3.5 | 4.7×10$^5$ | 0.85 | -0.87 | 1.2 | 224 |
| N50E39 | 8.5×10$^9$ | 0.4 | 5.2×10$^5$ | 0.0023 | -0.0054 | 0.0059 | 202 |
| Resultant | | | | | | 11.0 | 42 |

## 4.2 CME Deflection and the SEP Event

The 2015 June 21 CME was associated with an SEP event and has been discussed in detail in Piersanti et al. (2017). Here we point out two key results. (1) The SEP event is a soft spectrum event, and (2) the unusually large delay in the SEP onset. Figure 12a shows the time profile of the proton intensity at several SEP channels. The onset seems to be around 04:05 UT, after the start of the Wind/WAVES type II burst. Taking into account of the propagation time of 10 MeV protons (~40 min), we can place the start of the SEP event to the time the type II burst was still in the RAD2 domain. Significant particle counts can be found only in the lowest three energy channels with a tiny enhancement in the 89.8-114 MeV GOES energy channel.  Figure 12b shows the 10-100 MeV fluence spectrum, which is a closely related to CME kinematics. Hard spectrum events attain high speeds close to the Sun, while the soft spectrum events attain high speeds at large distances from the Sun. The spectrum in Fig. 12b is clearly very soft (power-law index is -6.32), similar to SEP events associated with filament eruptions from outside active regions (Gopalswamy et al. 2015b; 2016).  An additional reason for soft spectrum is poor magnetic connectivity.

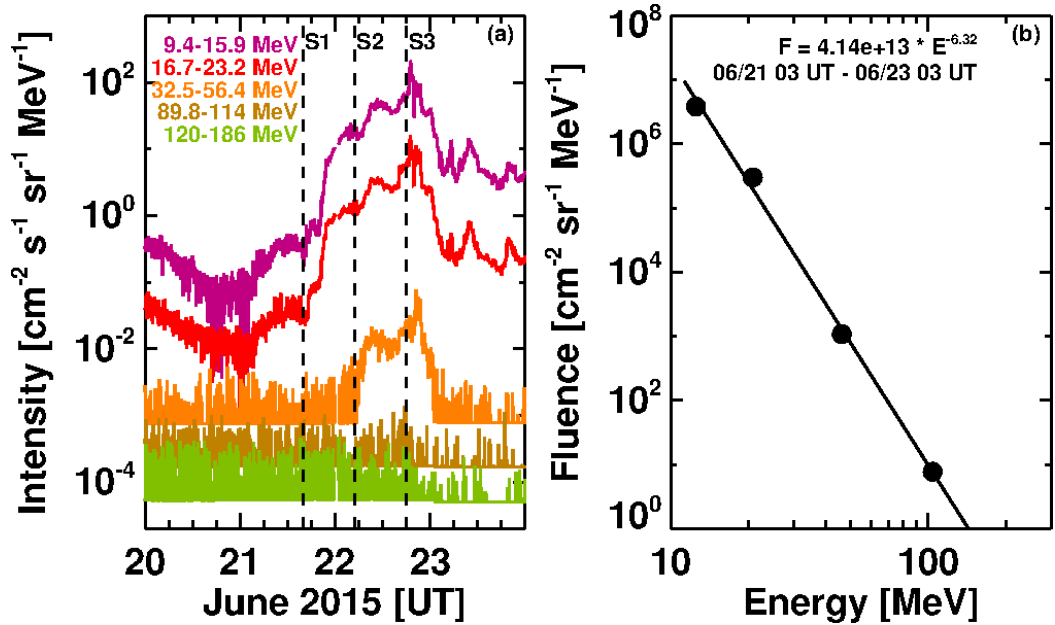

*Figure 12. Proton intensities and fluence spectrum of the 2015 June 21 SEP event. (a) The proton intensities are from GOES-13 observations in 5 energy channels. Significant proton intensity was found only in the lowest three channels, with a slight enhancement in the 89.8-114 MeV channel.  (b) The 10-100 MeV fluence spectrum of the SEP event. In computing the fluence, we used intensities in the time range 03:00 UT on June 21 to 03:00 UT on June 23. The background levels in the time range 18 UT on June 20 to 03:00 UT on June 21 were subtracted to obtain the spectrum. The spectrum can be fit to a power law in energy E, with an exponent of 6.32. The large exponent indicates a very soft spectrum, consistent with the slowly accelerating events.*

Ignoring the spike associated with the S1, we see the steady increase in SEP intensity around 19 UT on June 21. The SEP event became a large SEP event (according to NOAA scale of exceeding 10 pfu in the >10 MeV channel) only around 21 UT on June 21, right after the arrival of S2. Thus, only between the arrival times of S2 and S3 that the event can be considered as a large SEP event.  The unusually long delay of about 16.5 hours from the CME onset to the significant intensity increase is worth noting. An eastern source region in general should result in a gradual increase in SEP intensity at Earth due to gradual improvement of magnetic connectivity. In the present event, the source was barely eastern (E13) but the intensity increase was extremely gradual. We attribute the unusual delay to the poor connectivity due to the presence of the equatorial coronal hole (see Gopalswamy, 2012 for another event with a similar long delay). Movies of the eruption show that the rim of the coronal hole on the disk-center side brightened as the disturbance from the eruption approached it, indicating clear interaction with the coronal hole (https://sdo.gsfc.nasa.gov/assets/img/dailymov/2015/06/21/20150621_1024_0304.mp4). The brightening can be seen well at about 02:25 UT. The coronal hole is a high Alfven speed region, so the shock either weakens or ceases to be a shock before reaching the observer's field line until the shock moves far into the IP medium. It is likely that the CME deflection worsened the connectivity and hence delayed the increase in SEP intensity at Earth.

### 4.3 CME Deceleration

The empirical relation between the CME initial speed $u$ (km s$^{-1}$) and IP acceleration $a$ (m s$^{-2}$) is given by (Gopalswamy et al. 2001a),

(13)    $a = -0.0054(u - 406)$.

Using $u = 1780$ km s$^{-1}$ Eq. (13) gives $a = -7.4$ m s$^{-2}$. The IP acceleration determines the Sun-Earth travel time (*TT*) in the Empirical Shock Arrival (ESA) model (Gopalswamy et al. 2005b),

(14)    TT = $AB^u + C$ with $A=151.002$, $B=0.998625$, and $C=11.5981$.

Using $u = 1780$ km s$^{-1}$ and noting the source location N12E13 to get the earthward speed as 1697 km s$^{-1}$, we see that $TT = 26.2$ h. This is about 13 hours too soon compared to observations (39.5 h). One possibility is the effect of CME deflection (Gopalswamy et al. 2013a). Using the flux rope direction N11E25 reduces the earthward speed to 1584 km s$^{-1}$, and hence increases the travel time by 2.5 h to 28.7 h, which is still not adequate. This is probably because the single-view GCS fit does not give the full extent of deflection. Furthermore, the GCS fitting procedure has inherent uncertainties in the flux rope longitude (±17º) and latitude (±4º) as reported in Thernisien et al. (2009). Considering these uncertainties, it is not unreasonable that the derived GCS flux rope direction N11E25 could be N15E42. Such direction would give an earthward speed of 1278 km s$^{-1}$. The corresponding travel time is 37.6 h, which is close to the observed 39.5 h. In other words, the flux rope seems to have mainly deflected to the east by ~29º. It must be noted that such deflections of CMEs by coronal holes are very common (Gopalswamy 2015). The deflection also implies that the shock speed measured at Wind (776 km s$^{-1}$) corresponds to the shock flank and hence the nose speed is expected to be higher. As estimated above, the shock nose is about 42º away from the Sun-Earth line, giving a nose speed at 1-AU as 1044 km s$^{-1}$. This speed is consistent with the shock speed derived from type II burst data (see Table 2). Liu et al. (2015) used the sky-plane speed of the CME (~1300 km s$^{-1}$) to obtain the correct travel time. This is because the sky-plane speed is similar to the earthward speed after deflection. Both the earthward speed and the sky-plane speed are smaller than the true (deprojected) speed, so the sky-plane speed can often give better result than the deprojected speed does (similar to the earthward speed, see Gopalswamy et al. 2001c). However, the correct speed to use is the earthward speed as explained in Gopalswamy and Xie (2008).

**4.4 How to Reconcile the Metric and IP type II Bursts?**

Recall that the metric type II started at 02:24 UT at 180 MHz. From the height-time plot in Fig 6, we see that the shock nose is already at a height of 2.5 Rs. This height is too large for a starting frequency of 180 MHz. Gopalswamy et al. (2013b) obtained an empirical relation between the starting frequency and shock formation height as follows:

(15) $\quad f = 307.87 r^{-3.78} - 0.14$.

For $f = 180$ MHz, this relation gives $r = 1.2$ Rs, assuming fundamental emission. This is about half the heliocentric distance of the nose. Even if the emission were at the harmonic ($f = fp = 90$ MHz), the shock height given by Eq. (15) is only 1.4 Rs. This suggests that the type II emission must be coming from the flank of the CME, about 60° away from the nose. Furthermore, the nose speed of the shock is already 1300 km s$^{-1}$ by 02:24 UT (see Fig. 6). The flank speed is expected to be ~625 km s$^{-1}$. The Alfven speed is typically very low (~400 km s$^{-1}$) at a height of 1.2 Rs, whereas at 2.5 Rs it can be quite large, ~1000 km s$^{-1}$ (Mann et al. 2003; Gopalswamy et al. 2001b). Thus, the flank part of the shock has a higher Mach number than the nose part, consistent with the flank emission for the metric type II burst. The IP type II burst started at 02:36 UT, at which time the shock was at a height of 4.0 Rs and had attained a speed of 1700 km s$^{-1}$. At a height of 4 Rs, the Alfven speed decreases from the peak value, conducive for shock formation leading to the IP type II. For several minutes, there was an overlap between m type II emission from the flanks and IP type II emission from the nose, but the flank emission at metric wavelengths died off as the energy was mostly directed in the radial direction and given the larger Alfven speed in the flanks.

**4.5 Reconnected Flux and 1-AU Magnetic Field Strength**

Since the source region is well defined and the post eruption arcade (PEA) is clear during the CME in question, we were able to determine the total reconnected flux in the source region. Figure 13 shows the PEA and the underlying magnetic field distribution along with the compact dimming regions in which the legs of the CME flux rope are thought to be rooted (e.g., Webb et al. 2000). Clearly the northern dimming region is in the positive polarity region, while the southern one is in the negative polarity region. Therefore, the axial field of the flux rope points to the south, consistent with the field orientation from north to south in the filament and the high-inclination flux rope inferred from GCS fit to the CME.

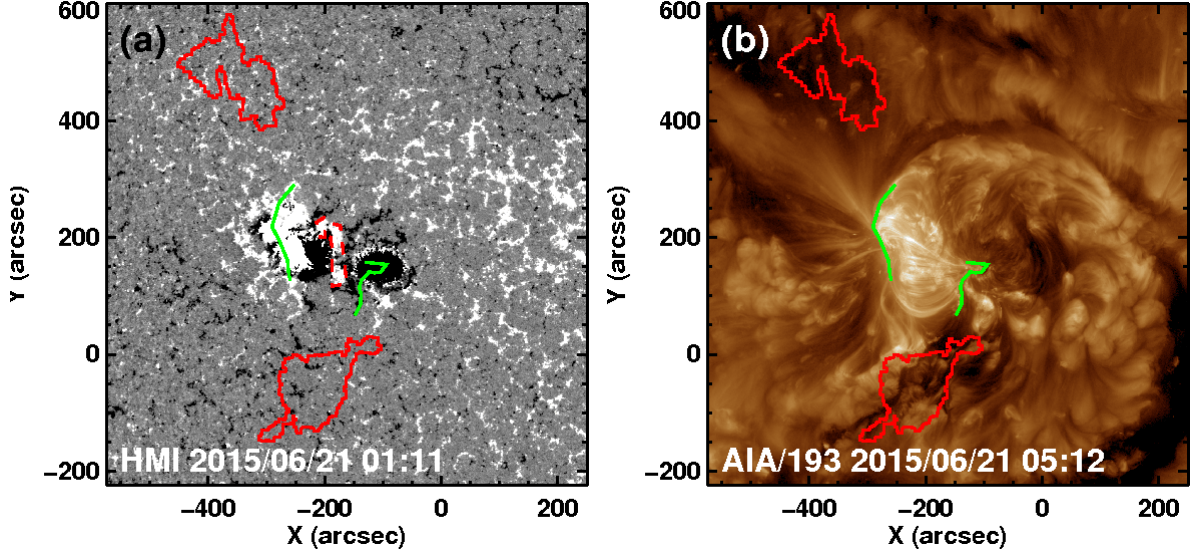

*Figure 13. Eruption configuration of the 2015 June 21 CME. Feet of the post eruption arcade (PEA, green lines) superposed on (a) a HMI line of sight magnetogram at 01:11 UT, and (b) an SDO/AIA image at 05:12 UT. In (a) the dark and white denote negative and positive magnetic polarities, respectively. The location of the twin-dimming is denoted by the red contours. The dimming regions are thought to be the locations where the CME flux rope is anchored to the photosphere. The poloidal flux of the CME flux rope is known to be approximately equal to the total reconnected flux in the eruption region. In (a), a small positive polarity region outlined by the dashed red contour is excluded in computing the magnetic flux underlying the PEA.*

We compute the total magnetic flux underlying the PEA area ($9.80 \times 10^{19}$ cm$^2$) that has an average unsigned field strength $<B>$ = 258.5 G as $2.53 \times 10^{22}$ Mx. Chertok et al. (2017) obtained a similar value for the total erupted flux ($2.34 \times 10^{22}$ Mx) although their criteria include the dimming flux, which is generally smaller than the arcade flux. According to the PEA technique developed by Gopalswamy et al. (2017a), the reconnected flux is half the flux under the PEA: $1.27 \times 10^{22}$ Mx. We can combine this information with the GCS flux rope fit to get the properties of the coronal flux rope using the FRED technique (Gopalswamy et al. 2017b,c). From the kappa value ($\kappa = 0.75$) obtained from the GCS fit, we get the minor radius of the flux rope ($R_0$) from the relation $R_0/R_{tip} = (1 + 1/\kappa)^{-1}$. At a height of 10 Rs for the shock, the flux rope $R_{tip}$ is ~9 Rs, giving the flux rope radius as 3.86 Rs. The axial field strength ($B_0$) of the coronal flux rope can be determined from the reconnected flux $\Phi_r$, the flux rope radius, and the flux rope length taken as 2 $R_{tip}$. i.e.,

(16) $\quad B_0 = \tfrac{1}{2} \Phi_r x_{01} / R_0 R_{tip}$ G,

where $x_{01}$ (=2.4048) is the first zero of the Bessel function $J_0$. Substituting $R_{tip}$ = 9 Rs, $R_0$ = 3.86 Rs, and $\Phi_r$ = $1.27 \times 10^{22}$ Mx, we get $B_0$ = 0.091 G (91 mG). This value is in the range of axial fields obtained for many coronal flux ropes obtained from the FRED technique (Gopalswamy et al. 2017b,c), but well above the average (31.5 mG). For a set of 37 CME-ICME pairs, flux ropes were fit independently at the Sun (at 10 Rs using FRED) and at 1 AU (using in-situ observations) to get an empirical relation between the axial magnetic fields as (Gopalswamy et al. 2017b):

(17)    $B_{0E}$ [nT] = 0.65$B_{0S}$ [mG] + 1.94.

Here $B_{0E}$ and $B_{0S}$ are the flux rope axial fields at Earth and at the Sun (10 Rs), respectively. Substituting $B_{0S}$ = 91 mG in Eq. (17), we get the expected $B_{0E}$ as 61.1 nT. The observed maximum and average field strengths at 1 AU were 25 nT and 21.3 nT, clearly much smaller than the expected axial field strength. Gopalswamy et al. (2017b) also obtained another empirical relation between $B_{0S}$ and the observed maximum field strength $B_{tE}$ at 1 AU as follows:

(18)    $B_{tE}$ [nT] = 0.46$B_{0S}$ [mG] + 3.16.

Putting $B_{0S}$ = 91 mG in Eq. (18), we get $B_{tE}$ = 45.0 nT, again much larger than the observed 25 nT. Assuming self-similar expansion, the 1-AU radius of the flux rope is expected to be 0.386 AU. We can get the radius of the flux rope at 1 AU from the ICME speed and duration as follows. Multiplying the central ICME speed of 680 km s$^{-1}$ (average between the leading- and trailing-edge speeds of 740 km s$^{-1}$ and 620 km s$^{-1}$) by the ICME duration (10 h), we get the ICME thickness of 0.16 AU. If it were a head-on flux rope, the radius would 0.08 AU, which is smaller by a factor of almost 5 than the expected 1-AU radius. As we discussed in section 4.1, the smaller observed size and magnetic field strength are likely due to the deflection of the CME away from the Sun-Earth line. We must point out that the kappa value is unusually large that might have contributed to the unusually large flux rope size derived above.

## 5. Summary and Conclusions

We described the Sun-to-Earth evolution of the 2015 June 21 CME-driven shock using white light, EUV, and radio remote sensing and in-situ observations in the solar wind. We were able to construct the eruption configuration involving flux rope feet locations (transient dimming regions), the post eruption arcade, and the EUV shock disturbance surrounding the source region. We computed the total reconnected flux in the reconnection process that created the CME flux rope and the PEA. The flux rope was driving a shock early in the event as inferred from the metric type II burst. The continued propagation of the shock was quantified using Wind/WAVES radio observations. The Wind/WAVES type II radio burst data represent a novel input to the analysis, which allowed us to track the shock from the corona to the location of the Wind spacecraft. In spite of the two intervening shocks passing by the Wind spacecraft, we were able to measure the shock speed at several locations in the inner heliosphere so that the increase in speed in the corona and the subsequent decrease were tracked. The shock speed evolution derived from the radio observations was consistent with the local shock speed at the Wind spacecraft. The plasma density at the Wind spacecraft also matched with that derived from the local type II burst observed by Wind/WAVES. While the nature of the ICME resulting from the eruption is not clear, the observed southward field is consistent with the well-known empirical relation that connects the storm strength to the ICME speed and the strength of the southward magnetic field. This paper thus provides a complete picture of the Sun-to-Earth evolution of the eruption in comparison with other results reported in previous papers. The main conclusions of this paper are listed below.

1. Although the June 21, 2015 CME was associated with two M-class flares, the first flare was confined, and therefore not associated with the CME. We confirmed the confined nature from lack of any mass motion and nonthermal electrons propagating away from the Sun.

2. The CME was accelerating throughout the inner heliosphere, somewhat similar to CMEs associated with filament eruptions from outside active regions. This event thus represents a hybrid event involving an active region filament extending outside the active region as a quiescent filament.

3. The magnetic field in the filament pointed from north to south, consistent with the north-south axis inferred from the EUV dimming locations, and the GCS fit to white-light observations.

4. We found evidence that the CME was deflected in its early phase, mainly by an equatorial coronal hole. Although the deflection did not change the north-south orientation of the flux rope, it made the ICME to have a shorter-than-usual duration at 1 AU.

5. The axial magnetic field strength (~91 mG) of the coronal flux rope is within the range of field strengths known from cycle-23 flux ropes, but is above average.

6. The expected 1-AU magnetic field strength and the flux rope radius are larger than the corresponding observed values, consistent with the deflected propagation.

7. The presence of southward magnetic field throughout the ICME is consistent with the high-inclination flux rope observed at the Sun with south-pointing axis.

8. Combining the shock speed evolution in the inner corona and the emission frequencies of the metric and IP type II bursts, we conclude that the metric emission originated from the flanks of the shock located approximately at 60º from the nose.

9. The combined data set captured the evolution of the shock from Sun to Earth: the initial rapid increase, slow increase in the outer corona, rapid decline within ~100 Rs, and finally a slower decline.

10. The storm strength (-217 nT) predicted by the empirical relation Dst = -0.01$VB$z – 32 nT is in close agreement with the observed minimum Dst value (-204 nT) during the storm.

11. The SEP event attained storm level after an unusually long delay, which can be explained by the presence of the equatorial coronal hole that weakened the western flank of the shock and probably also worsened the shock connectivity to an Earth observer.

12. The SEP event has a very soft fluence spectrum, consistent with the behavior of a CME with positive acceleration throughout the coronagraph field of view and poor magnetic connectivity.

13. Even though several CMEs occurred before and after the June 21 CME, they were all well isolated and hence did not influence the storm strength. The only notable interaction is the propagation of shock S3 into the sheath of shock S2 resulting in higher frequency of the local type II burst at Wind.

**Competing interests:**

The authors have no competing interests to declare.

**Acknowledgments**

This work benefited from NASA's open data policy in using SOHO,Wind, SDO/AIA, SDO/HMI, and OMNI (Operating Missions as a Node on the Internet, (ftp://spdf.gsfc.nasa.gov/pub/data/omni/) data. NoRH is currently operated by the Nagoya University in cooperation with the International Consortium for the Continued Operation of the Nobeyama Radioheliograph (ICCON). The BBSO operation is supported by NJIT, US NSF AGS-1250818, and NASA NNX13AG14G grants, and the NST operation is partly supported by the Korea Astronomy and Space Science Institute and Seoul National University and by the strategic priority research program of CAS with Grant No. XDB09000000.  X-ray and SEP data were obtained from NOAA's Space Weather Prediction Center. The interplanetary shock data base is supported by the NASA Living with A Star Targeted Research and Technology program under Grant NNX13AI75G: "Corona-Solar Wind Energetic Particle Acceleration (C-SWEPA) Modules." Work was supported by NASA's LWS TR&T program. NT was supported by NSF grant AGS-1622377.